\magnification=\magstep1
\pretolerance 2000
\baselineskip=20pt

\catcode`@=11
\def\vfootnote#1{\insert\footins\bgroup\baselineskip=12pt
  \interlinepenalty\interfootnotelinepenalty
  \splittopskip\ht\strutbox 
  \splitmaxdepth\dp\strutbox \floatingpenalty\@MM
  \leftskip\z@skip \rightskip\z@skip \spaceskip\z@skip \xspaceskip\z@skip
  \textindent{#1}\footstrut\futurelet\next\fo@t}
\skip\footins 20pt plus4pt minus4pt
\def\footstrut{\vbox to2\splittopskip{}}
\catcode`@=12
\pageno=1
\def\folio{\ifnum\pageno=0\else\ifnum\pageno<0 \romannumeral-\pageno
\else\number\pageno \fi\fi}
\def\buildurel#1\under#2{\mathrel{\mathop{\kern0pt #2}\limits_{#1}}}

\vglue 24pt plus 12pt minus 12pt

\input amssym.def
\input amssym.tex

\pretolerance 2000
\baselineskip=20pt

\catcode`@=11
\def\vfootnote#1{\insert\footins\bgroup\baselineskip=12pt
  \interlinepenalty\interfootnotelinepenalty
  \splittopskip\ht\strutbox 
  \splitmaxdepth\dp\strutbox \floatingpenalty\@MM
  \leftskip\z@skip \rightskip\z@skip \spaceskip\z@skip \xspaceskip\z@skip
  \textindent{#1}\footstrut\futurelet\next\fo@t}
\skip\footins 20pt plus4pt minus4pt
\def\footstrut{\vbox to2\splittopskip{}}
\catcode`@=12
\def\folio{\ifnum\pageno=0\else\ifnum\pageno<0 \romannumeral-\pageno
\else\number\pageno \fi\fi}
\def\buildurel#1\under#2{\mathrel{\mathop{\kern0pt #2}\limits_{#1}}}

\vglue 24pt plus 12pt minus 12pt

\bigskip

\centerline {\bf Remark on the (Non)convergence of Ensemble Densities in
Dynamical Systems}
\medskip
\centerline {\bf S. Goldstein$^1$, J. L. Lebowitz$^1$ and Y. Sinai$^2$}
\bigskip

\noindent {\bf Abstract}

We consider a dynamical system with state space $M$, a smooth, compact
subset of some ${\Bbb R}^n$, and evolution given by $T_t$, $x_t = T_t
x$, $x \in M$; $T_t$ is invertible and the time $t$ may be discrete,
$t \in {\Bbb Z}$, $T_t = T^t$, or continuous, $t \in {\Bbb R}$.  Here
we show that starting with a continuous positive initial probability
density $\rho(x,0) > 0$, with respect to $dx$, the smooth volume
measure induced on $M$ by Lebesgue measure on ${\Bbb R}^n$, the
expectation value of $\log \rho(x,t)$, with respect to any stationary 
(i.e.\ time invariant) measure $\nu(dx)$, is linear in $t$, $\nu(\log
\rho(x,t)) = \nu(\log \rho(x,0)) + Kt$. $K$ depends only on $\nu$ and
vanishes when $\nu$ is absolutely continuous wrt $dx$.
\bigskip
\centerline{************}
\bigskip
The time evolution of probability measures on the phase space $M$ of a
dynamical system depends both on the character of the dynamics, assumed
here to be given by a one parameter group of invertible measurable
transformations $T_t$, and the nature of the initial measure.  Given a
probability measure $\mu_0$ on $M$ at time $0$, the evolved measure at time
$t$, $\mu_t$, is such that the expectation value of functions $\phi(x)$ is
given by
$$
\mu_t(\phi) = \int_M \phi(x) \mu_t(dx) = \int \phi(T_{t} x)
\mu_0(dx),\eqno(1)
$$
or in terms of measurable sets $A \subset M$
$$
\mu_t(A) = \mu_0(T_{-t} A)\eqno(2)
$$
where $T_{-t} A$ is the set of points $y$ such that $T_{t} y$ belongs to
$A$.

There will typically be many stationary measures $\nu$, $\nu_t(dx) =
\nu(dx)$, for the dynamical system. Some are concentrated on the set of
fixed points or periodic orbits.  There can also be stationary measures
concentrated on fractal sets of zero Lebesgue measure.  This happens in
particular for generic Anosov systems; for other examples, see [1,2].  In
addition there may exist a stationary measure absolutely continuous with
respect to $dx$, i.e.\ $\nu(dx) = \bar \rho(x) dx$.  In the most familiar
examples of the latter situation $T_t$ preserves $dx$, as with Hamiltonian
flows on energy surfaces or the baker's transformation on the unit square,
in which case $\bar \rho(x)$ is constant, i.e.\ independent of $x$.

When $T_t$ is sufficiently ``chaotic'', the fractal and absolutely
continuous types of measures will generally have very good ergodic
properties, mixing or better.  For the case of an absolutely continuous
$\nu$, mixing implies that if $\mu_0(dx) = \rho(x,0)dx$ then
$$
\mu_t(f) = \int f(T_t x) \rho(x,0) dx = \int f(x) \rho(x,t) dx
\buildurel {t \to \pm \infty} \under 
\longrightarrow \int f(x) \bar \rho(x) dx
\eqno(3)
$$
for any bounded measurable $f(x)$.  The weak convergence of $\rho(x,t)$ to
$\bar \rho$, expressed by (3), is clearly compatible with the fact that
{\it when $T_t$ preserves $dx$}, the Gibbs entropy $S_\mu = -\int
\rho(x,t) \log \rho(x,t)dx$, and indeed any $\bar F = \int F(\rho(x,t))dx$, 
is constant in time.  

Unfortunately, it is sometimes thought that this constancy of $S_\mu$
for Hamiltonian evolutions is a manifestation of the conflict between
microscopic reversibility and the second law of thermodynamics, and
that the resolution of this conflict requires at least an acceptance
of weak convergence as the mathematical expression of the approach to
equilibrium characteristic of macroscopic irreversibility, and perhaps
even necessitates changes in the microscopic physical laws, c.f.\
[3b].  This concern and its proposed resolution are based on a
misunderstanding of the origin of the observed time asymmetry of
macroscopic physical systems, which really concerns not probability
densities but the behavior of individual systems whose microstates
$x_t = T_t x$ are points in a very high dimensional phase space
$M$. In fact, the second law refers not to $S_\mu$ but to an entropy
defined for individual macroscopic systems, whose observed
irreversible behavior is due first and foremost to the large
discrepancy between the scale of macroscopic observables (which behave
irreversibly) and microscopic scales and to the nature of ``typical''
initial conditions for the microstate $x$ of the system, c.f.\ [3].

These conceptual issues are, however, not the main concern of this
brief note, even though it was motivated by the paper of R. Fox in
this issue [4] in which such problems are discussed for the baker's
transformation.  In that paper Fox notes the constancy of $\bar F$
when $F = \log \rho$ for this transformation.  Here we are concerned
with what happens to functions of $\rho(x,t)$ when $T_t$ does not
preserve $dx$ and $\nu$ may not be absolutely continuous.

Let $\nu$ be a stationary probability measure, and let $\mu^{(1)}_t$
and $\mu^{(2)}_t$ be two measures on $M$, evolving according to the
dynamics.  If $\mu^{(1)}_t$ is absolutely continuous wrt to
$\mu^{(2)}_t$, i.e.\ $\mu^{(1)}_t(dx) = g(x,t)\mu^{(2)}_t(dx)$, then
it follows directly from (2) that
$$
g(x,t) = g(T_{-t}x, 0).\eqno(4)
$$
Suppose that $g(x,0)$ is continuous in $x$.  Then, given a function of
$g$, $f(g)$, integrable wrt $\nu$, we have $\nu(f(g(x,t))) =
\nu(f(g(x,0)))$ for all $t$.  Assume now that  $\mu^{(1)}_t$
and $\mu^{(2)}_t$ are themselves absolutely continuous wrt $dx$, with
continuous positive densities $\rho_1(x,t)$ and $\rho_2(x,t)$.  Then
$g(x,t) =
\rho_1(x,t)/\rho_2(x,t)$ and 
$$
\nu(f(g)) = \int_M f(\rho_1(x,t)/\rho_2(x,t)) \nu(dx) = Const.
\eqno(5)
$$

Setting $f(g) = \log g$ yields 
$$
\nu(\log \rho_1(x,t)) - \nu(\log \rho_2(x,t)) = C\eqno(6)
$$
independent of $t$.  Put now
$\rho_2(x,t) = \rho(x,t)$ and $\rho_1(x,t) = \rho(x,t+\tau)$. Eq. (5) then 
becomes for all $\tau$
$$
\nu(\log \rho(x,t+\tau)) - \nu(\log \rho(x,t)) = K(\tau).\eqno(7)
$$
Noting that $K(\tau_1 + \tau_2) = K(\tau_1) + K(\tau_2)$  
we obtain a rather surprising result 
$$
\nu(\log \rho(x,\tau)) = \nu(\log \rho(x,0)) + K\tau
\eqno(8)
$$
with $K$ independent of $\tau$.  In other words, the average of the
log of the density wrt the stationary measure $\nu$ is linear in the
time.  On the other hand it follows from (6) that the growth rate of
$\nu(\log
\rho(x,t))$ does not depend on $\rho$.  Hence $K$ depends
only on the dynamics $T_t$ and the stationary probability measure
$\nu$.  Consequently, we can compute $K$ by taking for our initial
(unnormalized) density $\rho(x,0) = 1$.  We then get
$$
K = \nu({d J_t \over dt}|_{t=0}),\eqno(9)
$$
where $J(x,t)$ is the Jacobian of the transformation $T_{-t}$, for
continuous time and 
$$
K = \nu(\log J(x))\eqno(10)
$$
where $J(x) = J(x,1)$, for discrete time.
If $\nu$ is absolutely continuous wrt $dx$, i.e.\ 
$\nu(dx) = \bar \rho(x)dx$, then putting $\rho_2(x,t) =
\bar \rho(x)$ and $\rho_1 = \rho$  in (5)  we see that 
$\int_M [\log\rho(x,t)] \bar \rho(x) dx$ is independent of $t$, i.e.,
$K$ vanishes for such a $\nu$.

In the case of a continuous time evolution given by a (smooth) vector
field, $\dot x = {\bf v}(x)$, the right side of (9) is just
$\nu(-\nabla \cdot {\bf v})$.  Eqs. (8) and (9) can then also be
obtained directly for a smooth, positive $\rho(x,0)$ by starting with
the continuity equation
$$
{\partial \rho(x,t) \over \partial t} = -\nabla \cdot (\rho{\bf v}(x)).
\eqno(11)
$$
We then find
$$\eqalign{
K &= {d \over dt} \int_M \log \rho(x,t) \nu(dx) = -\int_M \rho^{-1} \nabla
\cdot (\rho {\bf v})\nu(dx)\cr
~~~&=-\int_M [\nabla \cdot {\bf v} + (\nabla \log \rho)\cdot{\bf
v}]\nu(dx)}. \eqno(12)
$$
On the other hand, the time derivative of $\mu_t (\phi)$ is, for any
smooth $\phi(x)$, given by
$$
{d \over dt} \mu_t(\phi) = -\mu_t({\bf v} \cdot \nabla \phi).\eqno(13)
$$
Hence, by the stationarity of $\nu$, $\nu({\bf v} \cdot \nabla \phi) =
0$ and so the second term in the square bracket in (12) vanishes,
yielding explicitly
$$
K = -\nu(\nabla \cdot {\bf v}).\eqno(14)
$$

Eqs. (7) and (13) are to be compared with what happens to the rate of
change of the Gibbs entropy $S_\mu$, for $\mu_t(dx) = \rho(x,t)dx$.  A
straightforward computation gives 
$$
{d \over dt} S_\mu = -{d \over dt} \int \rho(x,t)\log \rho(x,t)dx =
\int_M (\nabla\cdot{\bf v})\rho(x,t)dx = \mu_t(\nabla \cdot {\bf v}).\eqno(15)
$$
$\dot S_\mu$ has been of much interest recently in connection with
``thermostatted'' nonequilibrium systems [1,2,5].  Under suitable conditions
on $T_t$, it can be shown that $\mu_t(dx) \buildurel {t \to \pm \infty}
\under \longrightarrow \nu_{\pm}(dx)$ with $\nu_+$ an SRB measure [1,2].
In such cases 
$$
-{d \over dt} S_\mu \buildurel {t \to \pm \infty} \under \longrightarrow
-\nu_\pm(\nabla \cdot {\bf v})
\eqno(16)
$$
with $\nu_+(\nabla \cdot {\bf v}) \leq 0$.  The equality holds if and only
if $\nu_+$ is absolutely continuous wrt $dx$, i.e.\ $\nu_+ = \bar \rho_+(x)
dx$.  On the other hand when $T_t$ is ``time reversible'' in the sense
that there exists a transformation $R$ on $M$, preserving $dx$, such that
$R^2 = I$ and $RT_t x = T_{-t} Rx$, then [1,2]
$$
K_+ = -\nu_+(\nabla \cdot {\bf v}) = \nu_-(\nabla \cdot {\bf v}) = -K_-.
\eqno(17)
$$
Thus, writing $S_{\pm}(t) = -\nu_\pm(\log \rho(x,t))$ we have in this case
that
$$
S_\mu (t) \sim S_\pm(t) \quad {\rm for} \quad t \to \pm \infty
$$
with
$$
S_\pm (t) = S_\pm(0) \mp K_+ t.
$$

As an illustrative example consider a flow on a circle, with $v(x) =
-\sin x + \omega$.  Here $x \in [-\pi,\pi]$ with periodic boundary
conditions and $\omega$ is a constant.  This example corresponds to a
particle moving in the plane with velocity $\bf u$ under the action of
an electric field ${\bf E}$ and a magnetic field $h$ perpendicular to
the plane.  The speed $|{\bf u}|$ is kept equal to one by a Gaussian
thermostat [2,5]; $x$ is the angle between the velocity ${\bf u}$ and
${\bf E}$ and $\omega \sim h/|E|$.  (This flow is time reversible,
with $R$ given by reflection through $\pi/2$, the minimum of $v$.)
For $|\omega| < 1$, $\nu_{\pm}(dx)$ are delta functions at
$x_\pm=\arcsin \omega$ with $|x_+| < |x_-|$.  We clearly have $K_+ =
\cos x_+ = \sqrt {1-\omega^2} = -K_- > 0$.  For $|\omega|> 1$ there is
a unique stationary state, $\nu(dx) = \bar \rho(x)dx$, with $\bar
\rho(x)$ proportional to $1/|v(x)|$ and $K=0$ on general grounds as
well as by explicit computation.  At $|\omega| = 1$, $x_+ = x_-$,
$\nu_+ = \nu_- =
\nu$ with $K=0$ so $K$ is continuous in $\omega$.  

Another observation which follows from (5) is that for an absolutely
continuous $\nu$, with density $\bar \rho(x)$,
$$
B_p = \int_M |{\rho(x,t) \over \bar \rho(x)} - 1|^p \bar \rho(x) dx
\eqno(18)
$$
is independent of $t$.  For $p=1$ (17) is just the $L_1$ distance between
$\mu_t$ and $\nu$; since $M$ is compact, $\int_M dx = |M| < \infty$, (17)
also implies, by the Schwartz inequality, that $\int_M |\rho(x,t) - \bar
\rho(x)|^2 dx \geq |M|^{-1} B_1^2 > 0$ unless $\rho = \bar \rho$, and a
similar statement is true of the higher norms.  Thus there can be no
convergence to zero of the $L_2$ and higher norms of $\rho(x,t) - \bar
\rho(x)$.

We conclude by noting that the long time behavior of $\rho(x,t)$ was
discussed in [6] for hyperbolic maps.  It was explained there that
conditional probability densities induced by $\mu_t$ on the unstable
manifolds converge, as $t \to \infty$, pointwise with their
derivatives to the corresponding densities given by $\nu$.  Along
stable directions, however, the densities $\rho(x,t)$ are extremely
irregular, as might be suggested by the preservation of the integrals
discussed above.

$^1$ Department of Mathematics and Physics, Rutgers University,
Piscataway, New Jersey.  oldstein@math.rutgers.edu, supported in part
by NSF Grant DMS 95--04556, lebowitz@math.rutgers.edu, supported in
part by NSF Grant DMR 95--23266

$^2$  Department of Mathematics, Princeton University, Princeton, New
Jersey, sinai@math.princeton.edu

{\bf References}

\item {[1]}  G. Gallavotti, Topics in Chaotic Dynamics, LNP, Springer, {\bf
448}, 271--311 (1995).  P. Garrido and J. Marro, eds.; D. Ruelle,
Dynamical Systems Approach to Nonequilibrium Statistical Mechanics: An
Introduction, IHES/Rutgers, Lecture Notes, 1997.

\item {[2]}  N. Chernov, G. Eyink, J. L. Lebowitz and Y. Sinai,
Steady-State Electrical Conduction in the Periodic Lorentz Gas,
{\it Commun. Math. Phys.} {\bf 54}, 569--601, 1993, and 
Derivation of Ohm's
Law in a Determinisitic Mechanical Model, {\it Phys. Rev.Lett.} {\bf 70},
2209--2212, 1993.

\item {[3a]}  J. L. Lebowitz, Boltzmann's Entropy and Time'Arrow, {\it
Physics Today}, {\bf 46}:32--38, 1993. 

\item {[3b]}  Responses to J. L. Lebowitz's article and his reply, {\it
Physics Today}, {\bf 47}:113--116, 1994.

\item {[3c]}  J. L. Lebowitz, Microscopic Reversibility and Macroscopic
Behavior: Physical Explanations and Mathematical Derivations, in Lecture
Notes in Physics, Springer (1994), J. J. Brey et al., eds.

\item {[4]}  R. Fox, Chaos, 

\item {[5]}  D. Evans and G. Morriss, Statistical Mechanics of
Nonequilibrium Liquids, Academic Press, 1990; W. G. Hoover, {\it
Computational Statistical Mechanics}, Elsevier, Amsterdam, (1991).

\item {[6]} Y. Sinai, {\it Topics in Ergodic Theory}, Lecture 18,
Princeton University Press, 1994.

\end